\documentclass[%
 reprint,
 amsmath,amssymb,
 aps,
]{revtex4-1}
\usepackage{color,soul}
\usepackage{graphicx}
\usepackage{dcolumn}
\usepackage{bm}

\usepackage{dsfont}
\usepackage{braket}
\begin{document}

\preprint{APS/123-QED}

\title{Simple embedding scheme for the spectral properties of the single-impurity Anderson model}

\author{Soumyodipto Mukherjee}
\author{David R. Reichman}%
\affiliation{
 Department of Chemistry, Columbia University, New York, NY, 10027, USA
}

\begin{abstract}
In this work we outline a simple and numerically inexpensive approach to describe the spectral features of the single-impurity Anderson model.
The method combines aspects of the density matrix embedding theory (DMET) approach with a spectral broadening approach inspired by
those used in numerical renormalization group (NRG) methods.  At zero temperature for a wide range of $U$, the spectral function 
produced by this approach is found to be in good agreement with general expectations as well as more advanced and complex numerical
methods such as DMRG-based schemes.  The theory developed here is simply transferable to more complex impurity problems.
\end{abstract}

\maketitle


\section{Introduction}

Quantum impurity models play a major role in modern condensed matter
physics.  As proxies of physical reality, impurity models encapsulate complex
physical behavior within the simple framework of a small subsystem
hybridized with an otherwise noninteracting bath.  Such models have laid
the foundation for our understanding of qualitative phenomena such as quantum 
dissipation, namely the generic features of how a bath induces dephasing and relaxation
in a small subsystem, as well as detailed specific and quantitative phenomena such as the description
of magnetic impurities in metals and the relaxation of tunneling centers in low-temperature
dielectric media\cite{Weiss}.  Due to central role played by impurity problems in condensed matter
physics, the development of methods for the the accurate description of their properties, in particular
their real-time or real-frequency behavior, remains at the forefront of the field.

One of the most important and well-studied impurity models is the Anderson model\cite{Anderson1961}.  The Anderson
Hamiltonian describes an electron-correlated quantum dot hybridized with
a non-interacting bath of fermions.  The physics of the Anderson model subsumes that of
the Kondo model, and thus describes the physics of dilute magnetic impurities in metals,
including the appearance of a resistivity minimum and the eventual formation of a 
$T=0$ singlet state induced by the complete
screening of the impurity spin by the conduction electrons\cite{Hewson}.
In the context of strongly correlated solids, the Anderson model has taken on renewed importance
with the development of dynamical mean-field theory (DMFT)\cite{dmft1, dmft2, dmft_review}, a 
quantum embedding approach where Hubbard-like models\cite{hub, hub_gutz, kana} are self-consistently
mapped to Anderson-like impurity problems.  In this latter regard,
the Anderson model and its variants provide a window into the 
properties of wide range of strongly correlated materials.

DMFT is a Green's function-based approach which, for many applications, requires the solution of
the spectral function of the Anderson model on the real-frequency axis. Despite the
reduction of complexity afforded by the mapping from the Hubbard model to the Anderson
model, the determination of real-time or real-frequency spectral behavior is a difficult task.
Simple quasi-analytical approaches like the non-crossing approximation\cite{nca3,nca2, nca1} are often not
sufficiently accurate in the most interesting physical regimes\cite{nca4}.
Analytic continuation of exact imaginary-time quantum Monte Carlo data is an ill-conditioned
numerical problem which can often produce artifacts\cite{JARRELL1996133, ctqmc1, ctqmc2, ctqmc3}, while exact diagonalization is
restricted to a small number of bath states, rendering a detailed accurate description of
features such as the high frequency behavior of the spectral function difficult\cite{dmft_ed, ctqmc3}.
Quantum chemistry methods such as truncated configuration interaction (CI) and complete active space configuration interaction (CAS-CI) have been employed 
as impurity solvers in DMFT\cite{dmft_ci}. Their accuracy is comparable to exact diagonalization solvers at a much lower cost. However, they 
share similar drawbacks due to the finite number of orbitals used in the CI expansion.


Recently great progress has been made in the formulation and use of time dependent 
density matrix renormalization group (DMRG) and related approaches\cite{td_dmrg, Verstraete2008}, as well
as sophisticated renormalization group approaches to study the behavior of the Anderson model
on the real time and frequency axis\cite{Peters2011, frg1, frg2, Ganahl2014, Chebyshev2}.  
These matrix product-based methods have been used as impurity solvers in one and multi-orbital 
DMFT calculations to obtain spectral functions of the one and two-band Hubbard models 
\cite{mps_dmft1, mps_dmft2, mps_dmft3}. Such techniques, while powerful, are also theoretically and numerically
involved.

In this work we propose a different and extremely simple method for the calculation of
the real-frequency spectral function of the Anderson model.  The approach combines the use
a density-matrix embedding theory (DMET) -like decomposition without self consistency\cite{Knizia2012, Knizia2013, Bulik2014_prb, Bulik2014_jcp,Wouters2016, Booth2015} with
a simple and physical protocol motivated by some implementations of the 
numerical renormalization group (NRG) for assigning line widths\cite{florens,Peters2011}.
The results are comparable to those produced by far more sophisticated theoretically 
and numerically involved approaches.  In principle the method outlined here is extendable
to more complex situations such as those that arise in cluster DMFT where
the embedded impurity-like system contains multiple sites or orbitals.

The paper is organized as follows: In Sec.\ref{sec:background} we introduce the single-impurity Anderson model (SIAM)
and briefly discuss the ground state embedding approach and the calculation of $T=0$ spectral functions. 
We propose a way to obtain the peak spectrum and a frequency-dependent broadening scheme is employed that generates smooth spectral functions. 
Sec.\ref{subsec:finite_siam} compares the ground state and spectral properties of a finite-sized SIAM to numerically exact results.
In Sec.\ref{subsec:thermo_lim} we analyze the spectral properties for the bulk case. Sec.\ref{conclusion} is devoted to our concluding remarks. 

\section{Background and Methodology}\label{sec:background}
\subsection{Model and embedding approach}\label{subsec:siam_gs}
The single-impurity Anderson model (SIAM)\cite{Anderson1961} describes a single interacting impurity coupled to a host conduction band of noninteracting fermionic levels. The Hamiltonian reads
\begin{eqnarray}
 \hat{H} &=& \hat{H}_{\text{cb}} + \hat{H}_{\text{imp}} + \hat{H}_{\text{hyb}} \nonumber \\
 &=& \sum\limits_{k\sigma} \epsilon_{k} \hat{n}_{k\sigma} + \sum\limits_{\sigma} \left( \epsilon_d + \frac{U}{2}\hat{n}_{d-\sigma}\right)\hat{n}_{d\sigma} \nonumber \\
 & &+ \sum\limits_{k\sigma} V_{k} (c^{\dagger}_{d\sigma}c_{k\sigma} + c^{\dagger}_{k\sigma}c_{d\sigma})~, \label{eq:siam_ham}
 \end{eqnarray}
where $\epsilon_k$ denotes the energy levels of the conduction band, $\epsilon_d$ the impurity level, $U$ the interaction on the impurity, $\hat{n}_{d\sigma} = c^{\dagger}_{d\sigma}c_{d\sigma}$ is 
the occupancy of the impurity, $\hat{n}_{k\sigma} = c^{\dagger}_{k\sigma}c_{k\sigma}$ is the band occupancy and $V_{k}$ the hybridization between the impurity and the conduction band. 
At this stage we defer the discussion of the choice of the hybridization parameter until later to keep the discussion general.

The approach we employ is based on a simplified version of the density matrix embedding theory (DMET) and its spectral extension\cite{Knizia2012, Knizia2013, Bulik2014_prb, Bulik2014_jcp,Wouters2016, Booth2015}. 
At the heart of the embedding procedure lies the Schmidt decomposition of the wave function. Formally, the Schmidt decomposition of a state is given by
\begin{equation}
 \label{eq:sd}
 \ket{\Psi} = \sum\limits_{i,j=1}^{m_{\alpha}} \lambda_{ij} \ket{\alpha_i}\ket{\beta_j}~,
\end{equation}
where the states $\ket{\alpha_i}$ represent the states that span the part of the system of interest, the fragment and the states $\ket{\beta_j}$ represent the states that span the rest of the system, the bath. $m_{\alpha}$ 
is the dimension of the fragment space assumed smaller than the bath. The Schmidt decomposition renders the wavefunction expansion in a compact form. Even for a complex state, the number of many-body states, $\ket{\beta_j}$, required to exactly define 
$\ket{\Psi}$, depends only on the size of the fragment which is generally much smaller than the bath. The projector $\hat{\mathcal{P}}=\sum_{ij}\ket{\alpha_i\beta_j}\bra{\alpha_i\beta_j}$ projects the full Hamiltonian onto the basis obtained from the Schmidt decomposition, 
$\hat{H}_{\text{emb}} = \hat{\mathcal{P}}\hat{H}\hat{\mathcal{P}}$. If $\ket{\Psi}$ were the exact ground state of the system, $\hat{H}_{\text{emb}}$, a much smaller Hamiltonian, would yield the exact ground-state energy. 
However, this procedure is purely formal as we would need to know the exact ground-state wavefunction from the outset, which is impossible to obtain for large correlated systems.

The mean-field solution for the ground state is the single Slater determinant $\ket{\Phi^{(0)}}$. It's Schmidt decomposition can be obtained from single-particle linear algebra\cite{Peschel2009,Peschel2012} at a cost no greater than 
the diagonalization of the single-particle Hamiltonian. It has the same form as Eq.~\ref{eq:sd}, where the many-body fragment state $\ket{\alpha_i}$ is constructed from the single-particle states corresponding to the fragment 
sites (fragment orbitals). For the case of a Slater determinant, the many-body bath states $\ket{\beta_j}$ take a particularly simple form\cite{Knizia2013}. In particular, they are constructed from a single-particle bath state (bath orbital) multiplied by a determinant of core electrons.  
These core orbitals have no overlap with the fragment space. The details of constructing the single-particle fragment, bath and core orbitals are outlined in Refs. \cite{Knizia2012, Knizia2013, Bulik2014_prb, Bulik2014_jcp,Wouters2016}.

The ground-state embedding approach for SIAM is summarized as follows:
\begin{enumerate}
 \item First, a restricted Hartree-Fock (RHF) calculation which yields a single Slater determinant, $\ket{\Phi^{(0)}}$, is used as an approximate ground-state of the SIAM Hamiltonian.
 \item A set of local fragment site(s), which includes the interacting impurity, is chosen. The Schmidt decomposition of $\ket{\Phi^{(0)}}$ is performed. Importantly, this decomposition yields a single-particle basis which is used to construct the interacting embedding Hamiltonian, $\hat{H}_{\text{emb}}$. 
 This can be understood as a simple basis transformation of the full SIAM Hamiltonian from the single-particle fermionic site basis to the single-particle Schmidt basis.
 \item $\hat{H}_{\text{emb}}$ is easily diagonalized to obtain an approximate correlated ground state wavefunction, $\ket{\Psi_{\text{emb}}}$. All expectation values are calculated using 
 $\braket{\Psi_{\text{emb}}|\hat{O}|\Psi_{\text{emb}}}$.
\end{enumerate}
It is to be noted that $\ket{\Psi_{\text{emb}}}$ has the same form as the Schmidt decomposed $\ket{\Phi^{(0)}}$, the only difference being that $\ket{\Psi_{\text{emb}}}$ is obtained after including the correlations on the fragment whereas the latter included them in a mean-field way. 


For the SIAM, there is no self-consistency as the interactions are localized on the impurity and the bath states made out of the conduction band levels is, by default, noninteracting. 
At $T=0$, $\mu=0$ in the SIAM Hamiltonian in Eq.~\ref{eq:siam_ham} maintains half-filling on the impurity. Since the two-electron Coulomb term appears only on the impurity in $H_{\text{emb}}$, $\mu=0$ still ensures the required half-filling. 
In this respect the procedure outlined is a `single-shot' embedding analogous to the $u=0$ case discussed in \cite{Wouters2016}.

A translationally-invariant system like the Hubbard model can be divided into identical fragments each of which is embedded in its own bath. 
Once the self-consistency has been achieved, the fragment's contribution to the ground state energy is evaluated. 
The total ground state energy is a sum of such identical contributions from each of the fragments. For SIAM, a non-translationally-invariant system, we have only one fragment which includes the impurity site and its corresponding bath. The contribution of the unentangled core orbitals to the total ground state energy cannot be neglected in this case. 
Performing DMET with this unique partitioning also results in an approximate correlated wavefunction for the full system which in a `single-shot' gives variational estimates of the ground state energy.

\subsection{Dynamics}
 The quantum embedding approach has been generalized to dynamic properties, particularly for the evaluation of bulk spectral functions\cite{Booth2015, Chen2014}. Here, we apply the method to evaluate the single-particle, local density of states (LDOS) for the SIAM and provide extensions to the current formulation.
 The technique builds on the ground-state formalism where a bath space of frequency-dependent many-body states is constructed by the Schmidt decomposition of the linear response vector given by
 \begin{equation}
  \ket{\Phi^{(1)}(\omega, \eta_1)} = \frac{1}{\omega - (h - \epsilon_0) + i\eta_1}\hat{V}\ket{\Phi^{(0)}}~,
  \label{eq:lin_res_vec}
 \end{equation}
 where $h$ is the single-particle Hamiltonian with $\epsilon_0$ it's ground-state energy. $\hat{V}=c_d^{(\dagger)}$ is used for the evaluation of the impurity LDOS. $\ket{\Phi^{(1)}(\omega, \eta_1)}$ can be expressed in a Schmidt decomposed form \cite{Booth2015}
 \begin{equation}
  \ket{\Phi^{(1)}(\omega, \eta_1)} = \sum\limits_{ijm} \lambda_{ij} \hat{A}^{(m)}(\omega,\eta_1)\ket{\alpha_i}\hat{B}^{(m)}(\omega,\eta_1)\ket{\beta_j}~,
 \end{equation}
 where $\hat{A}^{(m)}(\omega,\eta_1)$ and $\hat{B}^{(m)}(\omega,\eta_1)$ are operators that act on the fragment and bath states, respectively. 
 This defines a frequency-dependent basis and the corresponding projector 
 $\ket{K_{ijm}(\omega, \eta_1)} = \ket{\alpha_i}\otimes\hat{B}^{(m)}(\omega,\eta_1)\ket{\beta_j}$, $\hat{\mathcal{P}} = \sum_{ijm}\ket{K_{ijm}(\omega, \eta_1)}\bra{K_{ijm}(\omega, \eta_1)}$.
 
Defining $\hat{H}'(\omega, \eta_1) = \hat{\mathcal{P}}\left(\hat{H} - E_{gs}\hat{\mathds{1}}\right)\hat{\mathcal{P}}$, where $E_{gs}$ is the energy corresponding to the frequency-independent ground state $\ket{\Psi_{\text{emb}}}$ discussed in the previous section. The approximate embedding Green's function is given by  
 \begin{equation}
  G(\omega, \eta_1, \eta_2) = \bra{\Psi_{\text{emb}}}\hat{X}'\frac{1}{\omega - \hat{H}'(\omega, \eta_1) + i\eta_2}\hat{V}'\ket{\Psi_{\text{emb}}}~,
  \label{eq:gf_pre_peak}
 \end{equation}
 where $X=V^{\dagger}$, $\hat{X}'=\hat{\mathcal{P}}\hat{X}\hat{\mathcal{P}}$ and $\hat{V}'=\hat{\mathcal{P}}\hat{V}\hat{\mathcal{P}}$ gives the single-particle impurity Green's function with it's imaginary part as the single-particle density of states, namely
 \begin{equation}
  A(\omega, \eta_1,\eta_2) = -\frac{1}{\pi}\Im[G(\omega,\eta_1,\eta_2)]~.
  \label{eq:sp_dos}
 \end{equation}
 This generates a Lorentzian broadened spectral function dependent on the parameters $\eta_1$ and $\eta_2$. A common choice is to set $\eta_1=\eta_2$\cite{Booth2015, Peters2011}. This choice might not always be desirable, especially for a discretized SIAM, where appropriate broadening of the peak spectrum 
 leads to better resolution of the spectral function at both high and low energies\cite{florens, Bulla2008, Zitko2009}. This emphasizes the need for a more detailed construction of spectral broadening, as discussed below.
 It should be noted that the formulation can be extended to more complex two-particle Green's functions with an appropriate choice for $\hat{V}$. For example, $\hat{V} = \sum_{\sigma} = c_{i\sigma}^{\dagger}c_{i\sigma}$ is used to evaluate the local 
 density-density response function\cite{Booth2015}.
 \subsection{Peak spectrum and broadening}\label{subsec:peak_broad}
 The size of the frequency-dependent Schmidt basis generally makes it manageable to perform full exact diagonalization (ED) on $\hat{H}'(\omega, \eta_1)$ giving access to its entire eigenspectrum. This allows one to express Eq.~\ref{eq:gf_pre_peak} in an explicit Lehmann representation.
 Since the functional form of the eigenvectors, $\ket{\Psi'_n(\omega,\eta_1)}$ and the eigenvalues $E'_n(\omega,\eta_1)$ of $\hat{H}'(\omega,\eta_1)$ are not known, $\eta_1$ is chosen to be very small to obtain results independent of it. 
 The value of $\eta_1$ determines the number of delta peaks observed when calculating the peak spectrum under the following guideline. Hence, for all the results in Sec.\ref{subsec:thermo_lim}, we have used $\eta_1=10^{-8}$ which ensured the convergence of the number of delta peaks.
 For a more detailed discussion on the behavior of $E'_n(\omega,\eta_1)$ for different values of $n$ and $\eta_1$, see Appendix~\ref{app: eta1}. The single-particle Green's function is given by 
 \begin{equation}
  \label{eq:lehmann}
  G(\omega,\eta_2) = \sum\limits_n\frac{|\braket{\Psi'_n(\omega)|c_d^{\dagger}|\Psi_{\text{emb}}}|^2}{\omega-E'_n(\omega)+i\eta_2}~,
 \end{equation}
 where we have used
 \begin{equation}
  \hat{H}'(\omega) = \sum\limits_n E'_n(\omega)\ket{\Psi'_n(\omega)}\bra{\Psi'_n(\omega)}~.
 \end{equation}
 If $\omega-E'_n(\omega)=0$ has $m$ solutions given by $\omega_m^{(n)}$, the denominator of the imaginary part of $G(\omega,\eta_2)$ in Eq.\ref{eq:lehmann} can be approximated using a Taylor series around $\omega_m^{(n)}$ up to second order in $\omega$. 
 Integrating over all frequencies gives the spectral function in terms of delta functions at the peak positions and it's corresponding weights,
 \begin{equation}
  A(\omega) = \sum\limits_{mn} A_{mn}\delta(\omega-\omega_m^{(n)})~,
  \label{eq:peak_pos}
 \end{equation}
 and the spectral weight is given by
 \begin{equation}
  A_{mn} = \frac{|C_n(\omega_m^{(n)})|^2}{\left|1-\frac{dE'_n\left(\omega_m^{(n)}\right)}{d\omega}\right|}~,
  \label{eq:spec_weight}
 \end{equation}
 where $|C_n(\omega)|^2 = |\braket{\Psi'_n(\omega)|c_d^{\dagger}|\Psi_{\text{emb}}}|^2$. 
 The peak spectrum is convoluted with a Gaussian kernel\cite{Peters2011, Bulla2008, Zitko2009}
 \begin{equation}
  K(\omega,\omega') = \frac{1}{b\sqrt{\pi}}\exp\left[-(\omega-\omega')^2/b^2\right]~.
 \end{equation}
 The width, $b$ is chosen to be frequency dependent \cite{florens,Peters2011} and of the form
 \begin{equation}
  b = c_1\omega_{\text{min}} + c_2|\omega'|~,
  \label{eq:width}
 \end{equation}
 where $\omega_{\text{min}}$ is the position of the lowest lying peak above the fermi energy, and $c_1$ and $c_2$ are positive constants. We delay our discussion on how to appropriately choose these constants until Sec.\ref{subsec:thermo_lim}. The smooth spectral function is then given by
 \begin{eqnarray}
  A(\omega) &=& \int\limits_{-\infty}^{\infty}K(\omega, \omega')A(\omega') \\
  & &\sum\limits_{mn}\frac{A_{mn}}{b\sqrt{\pi}}\exp\left[-(\omega-\omega_m^{(n)})^2/b^2(\omega_m^{(n)})\right]~,
 \end{eqnarray}
 with $A_{mn}$ obtained from Eq.\ref{eq:spec_weight}.
 \subsection{Comparison to numerical renormalization-group methods}
 Since the first applications to the SIAM\cite{Krishna-Murthy1980}, non-perturbative approaches like the numerical renormalization-group method(NRG)\cite{Wilson1975} have been successful in describing both static thermodynamic properties and dynamical response and spectral functions\cite{Bulla2008,Hewson}. 
 These approaches have seen the introduction of a variety of techniques aimed at improving accuracy and resolution in both the high and low frequency regions \cite{florens, Bulla2008}.
 NRG relies on a logarithmic discretization of the conduction band which maps the Hamiltonian of Eq.~\ref{eq:siam_ham} onto a chain geometry. The mapping is analytically performed with the energy levels of the conduction band arranged according to 
 $\epsilon_n = \pm D\Lambda^{-n}$, where $D$ is the conduction band edge and $\Lambda > 1$ is the discretization parameter. The mapping is exact in the limit $\Lambda \to 1$. For a conduction band with constant hybridization $V_{k}=V$ and a flat density of states 
 $\rho(\omega)=\sum_k\delta(\omega-\epsilon_k)=1/2D$ for $\omega \in [D,D]$, the mapping leads to the chain Hamiltonian\cite{Wilson1975}
 \begin{eqnarray}
  H &=& \sum\limits_{\sigma} \left( \epsilon_d + \frac{U}{2}\hat{n}_{d-\sigma}\right)\hat{n}_{d\sigma} + V\sum\limits_{\sigma}\left(c_{d\sigma}^{\dagger}c_{1\sigma} + \text{H.c} \right) \nonumber \\
  & &+\sum\limits_{\sigma,n=1}^{\infty}t_n\left(c_{n\sigma}^{\dagger}c_{n+1\sigma} + \text{H.c}\right)~, \label{eq:log-disc-ham}
 \end{eqnarray}
 with $\Gamma=\pi V^2\rho(0)$ and
 \begin{equation}
  t_n = \frac{D\Lambda^{-n/2}\left(1+\Lambda^{-1}\right)\left(1-\Lambda^{-n-1}\right)}{\left(2\sqrt{\left(1-\Lambda^{-2n-1}\right)\left(1-\Lambda^{-2n-3}\right)}\right)}~.
  \label{eq:log-hop}
 \end{equation}
 In NRG, the Hamiltonian in Eq.\ref{eq:log-disc-ham} is iteratively diagonalized increasing the number of sites from the impurity at each iteration and truncating the high-energy states 
 if the number of states in the Fock space exceeds a certain chosen limit.
 
 More recent methods such as DMRG and related approaches\cite{Schollwock2011, Verstraete2008} use a variational optimization to obtain the ground state, allowing for feedback from lower to higher energies, which are absent in the NRG approach. 
 DMRG-like methods allow for an arbitrary discretization of the conduction band. This has the advantage of improving the spectral resolution at high energies by incorporating a linear discretization instead of a logarithmic one which has fewer states at high energies.
 In NRG calculations, the discretization parameter ranges from $\Lambda=1.5$ to $2.5$\cite{Bulla2008}. Lower values of $\Lambda$ imply that more states are retained in each NRG iteration, making it computationally challenging with increasing number of iterations. 
 This is not the case for DMRG or MPS-based methods where values as low as $\Lambda=1.05$ have been used\cite{Ganahl2014}. These advantages of DMRG, however, come at the cost of the loss of direct access to the full spectrum of excited states.
 
 The embedding scheme proposed here is similar to DMRG or MPS-like methods in terms of the advantages it bears, but is far simpler. If a logarithmic discretization is employed, although not necessary, the infinite chain in Eq.\ref{eq:log-disc-ham} may cut-off at a finite $N$. The 
 process of obtaining the spectral functions starts with the evaluation of the embedded ground-state. In this work, we have used $N\sim300$. Such large values can be handled easily because the embedded system that ultimately needs to be fully diagonalized is independent of $N$, and only depends on the size of the fragment embedded.

 \section{Results}\label{sec:results}
\subsection{Finite-sized SIAM}\label{subsec:finite_siam}
  \begin{figure}[t]
\centering
\includegraphics[width=80mm,height=50mm]{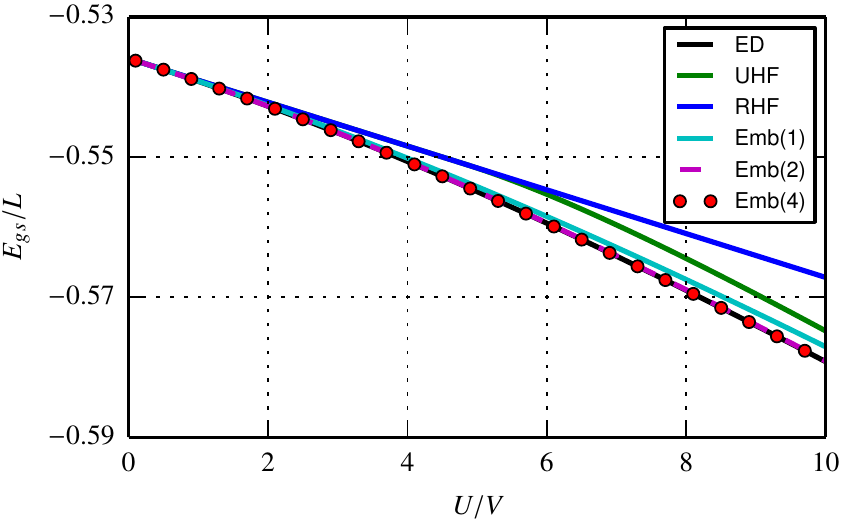}
\caption{\label{fig:siam_gs_energy} Ground-state energy per-site $E_{gs}/L$ $(L=8)$ as a function of $U$ for the symmetric SIAM from single-site and cluster embedding. $V_k=V=0.1$ and Emb(n) refers to the number of sites included in the fragment.}
\end{figure}
 In order to assess the performance of the prescribed embedding method for the SIAM, we first investigate a small-sized system and the results are compared against mean-field approaches like RHF and UHF and exact results obtained from Exact Diagonalization (ED). For this case, the SIAM is represented by an interacting impurity coupled to seven noninteracting conduction band levels (L=8)
at half-filling. We consider the symmetric case with $\epsilon_d = -U/2$. The seven conduction band levels are evenly spaced on $[-1,1]$. The hybridization energy from the impurity to the conduction band is taken to be same for all the conduction band levels with $V_k = V = 0.1$.
\begin{figure}[b]
\centering
\includegraphics[width=80mm,height=50mm]{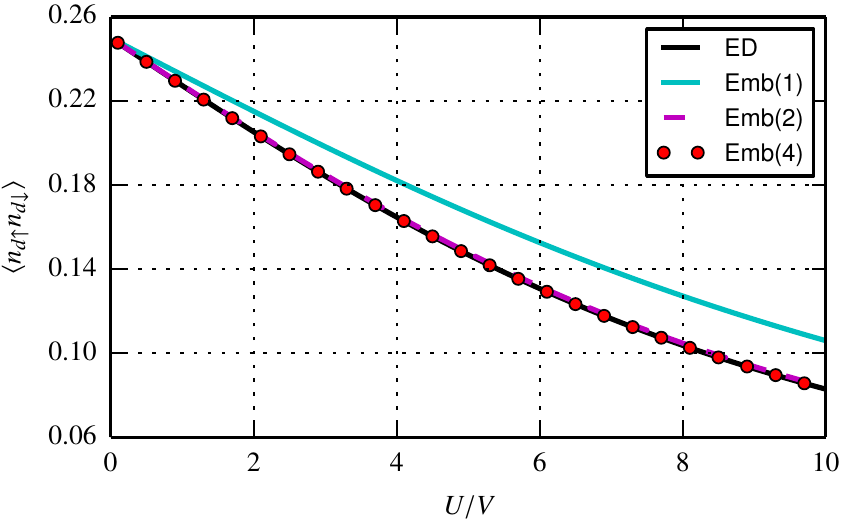}
\caption{\label{fig:siam_gs_docc} Double occupancy of the impurity $\braket{n_{d\uparrow}n_{d\downarrow}}$ for the symmetric SIAM with the same parameters as used in Fig. \ref{fig:siam_gs_energy}.}
\end{figure}
Fig.\ref{fig:siam_gs_energy} shows the ground-state energy per-site $E_{gs}/L$ as a function of $U$. The various levels of embedding depend on the number of sites included in the fragment. Emb($n$) refers to a fragment with $n$ sites which is coupled to a bath of the same size. The cost amounts to solving a $2n$ sized system which is achieved via ED for small $n$.
For $n=1$, only the impurity is part of the fragment. For higher values of $n$, the fragment consists of the impurity and the $n-1$ conduction band levels closest to the fermi energy. 

The embedding approach performs better than mean-field methods like RHF and UHF for the entire range of $U$. This is expected as the embedding improves over RHF by taking into account the interactions on the impurity explicitly. Even Emb(1) which is equivalent to solving a two-site system is significantly better than standard mean-field methods. Higher levels of 
embedding systematically improve the energies becoming numerically exact when $n=L/2$. This can be seen in Emb(4) which reproduces the ED result at the same cost as doing the ED calculation.

Fig.\ref{fig:siam_gs_docc} shows the double occupancy of the impurity $\braket{n_{d\uparrow}n_{d\downarrow}}$ as a function of $U$. Emb(1) does not capture the correct curvature due to the lack of short-range pairing in the small embedded system. 
However, cluster embedding generates systematic improvements over the single-site case. Emb(2) is visibly indistinguishable from the exact answer while Emb(4) is numerically exact, as expected.

\begin{figure}[t]
\centering
\includegraphics[width=85mm,height=65mm]{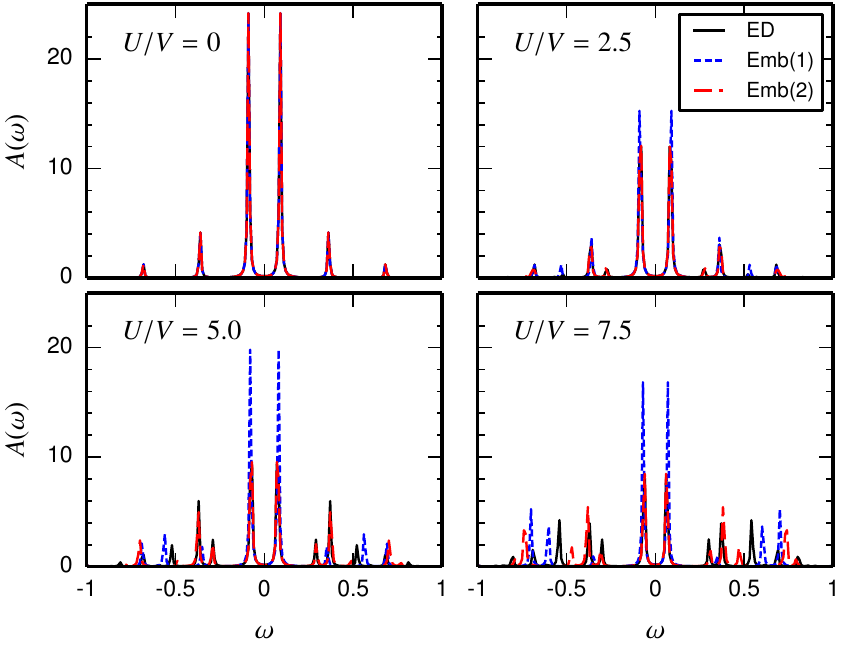}
\caption{\label{fig:siam_small_dos}  $T=0$ single-particle impurity spectral function $A(\omega)$ obtained from one and two-site embedding for the symmetric SIAM with $L=8$, $V_k=V=0.1$ and $\eta_1=\eta_2=0.005$ for different interaction strengths.}
\end{figure}

\begin{figure}[b]
\centering
\includegraphics[width=85mm,height=65mm]{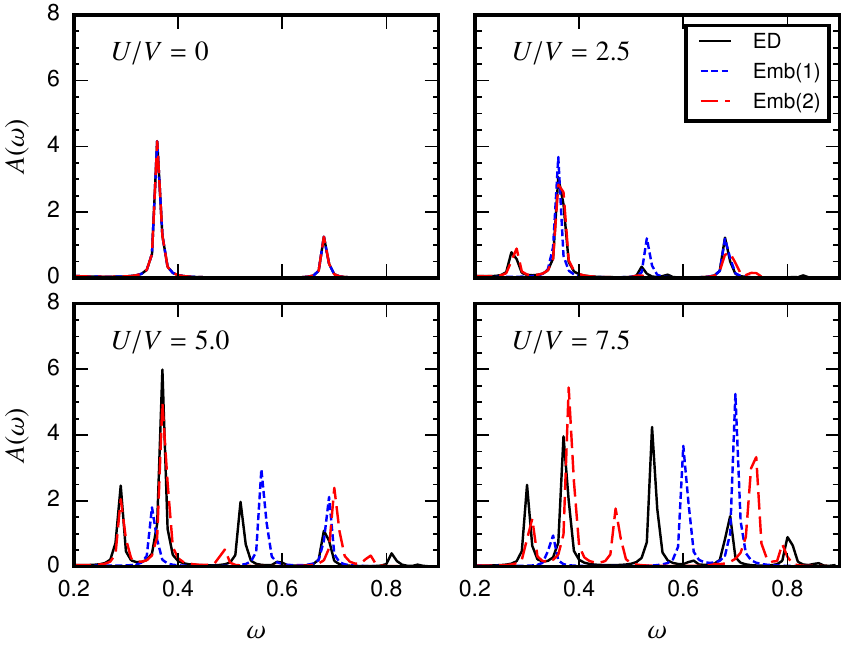}
\caption{\label{fig:siam_small_dos_mag}  Magnified view of Fig.~\ref{fig:siam_small_dos} showing details of the high-energy spectral features.}
\end{figure}

Fig.\ref{fig:siam_small_dos} compares the spectral functions obtained from different levels of embedding approximations with exact results from ED for varying strength of interaction, $U$. The Lorentzian broadened spectral functions are obtained from Eq.~\ref{eq:sp_dos} with $\eta_1=\eta_2=0.005$. 
The embedding results are exact in the noninteracting case, as expected, for both single-site and cluster embedding as seen in the top left panel of Fig.\ref{fig:siam_small_dos}. 
Both Emb(1) and Emb(2) correctly predict
the positions of the low energy excitations. As $U$ increases, the weight of excitations 
around the Fermi energy decreases and peaks at
higher energies are observed. Emb(2) displays this general trend whereas Emb(1) shows 
some oscillating behavior for the weight of these low energy excitations. Emb(2), as expected, is 
able to resolve features in the range $0\le\omega\le0.4$ better than Emb(1) as seen in Fig.~\ref{fig:siam_small_dos_mag}. 
The accuracy of Emb(2) is high at low values of $U$. At 
high values of $U$, particularly at $U/V=7.5$, Emb(2) is able to resolve the positions 
of some of the peaks in this range but does not quantitatively capture the weights of these excitations.


The systematic improvement with cluster embedding and its promising performance on small systems motivates the application of the technique to systems in the thermodynamic limit.

\subsection{Thermodynamic limit}\label{subsec:thermo_lim}
\begin{figure}[t]
\centering
\includegraphics[width=85mm,height=65mm]{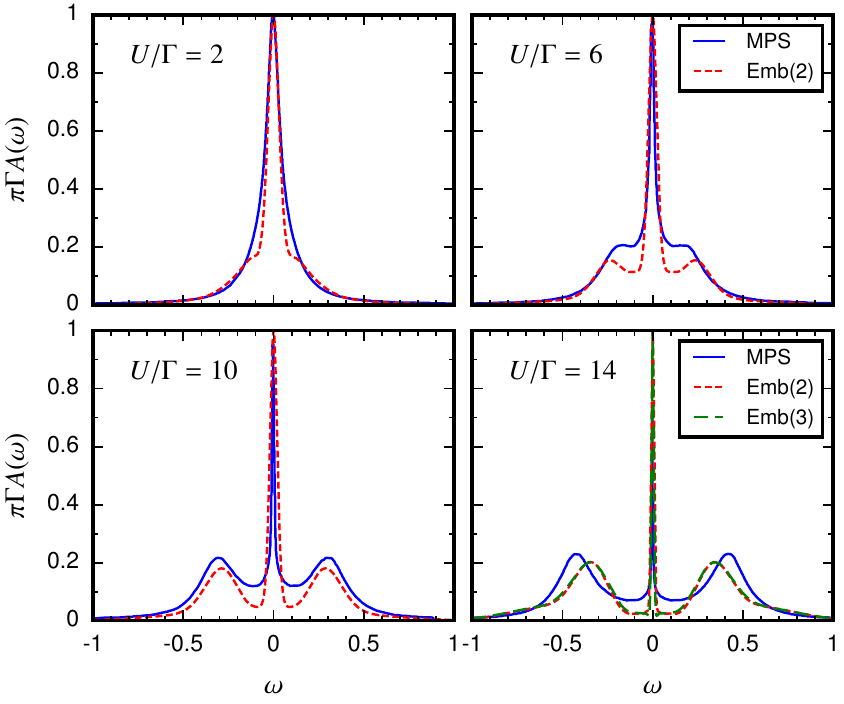}
\caption{\label{fig:siam_dos_log} $T=0$ impurity spectral function obtained from two-site embedding (Emb(2)) and Chebyshev-MPS for the symmetric SIAM with logarithmically discretized conduction band\cite{Ganahl2014}. The total number of sites in the chain, including the impurity, is $300$. $\Lambda=1.05$, $\Gamma=0.05$ and $D=1$. 
$c_1\approx 0.3-0.4$ and $c_2\approx 6.8-18.8$. The different panels correspond to different values of interaction. Bottom right panel also includes the spectral function obtained from three-site embedding (Emb(3)).}
\end{figure}
 For a single impurity coupled to a continuous conduction band, we take a conduction band with $D=1$ and use a logarithmic discretization with $299$ conduction band states distributed in the interval $[-D,D]$. The discretization parameter is taken as $\Lambda=1.05$ and the hybridization as $\Gamma=0.05$. Fig.~\ref{fig:siam_dos_log} 
 shows the $T=0$ impurity spectral function using two-site embedding in the wide-band limit $D\gg\Gamma$, where in each panel we have considered different values of the interaction strength from the weak to strong-coupling regime. Our results are compared to calculations using the Chebyshev-MPS method in the wide-band limit where the computed Chebyshev moments are post-processed with linear prediction\cite{Ganahl2014}. 
 These results achieve similar precision as the dynamical density matrix renormalization group method(DDMRG)\cite{Holzner2011} at a lower cost, and serves as a good benchmark for low-cost embedding methods. 

 \begin{figure}[t]
\centering
\includegraphics[width=85mm,height=65mm]{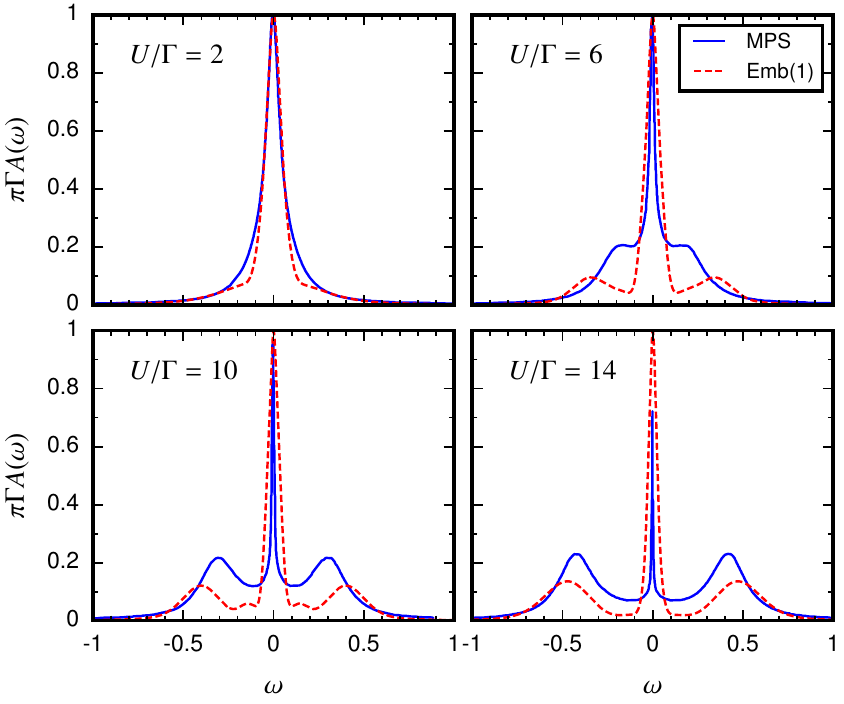}
\caption{\label{fig:one-site_dos} $T=0$ impurity spectral function obtained from single-site embedding (Emb(1)) for the symmetric SIAM with a linearly discretized conduction band. $299$ conduction band levels are evenly spaced in $\omega\in[-1,1]$. $\Gamma=0.05$ and the range of interactions considered is the same as in Fig.~\ref{fig:siam_dos_log}. 
$c_1\approx 0.3-0.35$ and $c_2\approx 6.1-12.0$.}
\end{figure}

 The delta peaks and their corresponding weights are obtained from Eq.~\ref{eq:peak_pos} and Eq.~\ref{eq:spec_weight} respectively. As discussed in Sec.~\ref{subsec:peak_broad}, the frequency-dependent Gaussian broadening kernel has two parameters, $c_1$ and $c_2$ (see Eq.~\ref{eq:width}). 
 At high values of $U$, the resolution of the Hubbard satellites is almost entirely governed by $c_2$ as $c_1$ primarily affects the low energy resolution. As a result, $c_2$ is chosen such that for strong interactions the high-energy Hubbard satellites have a 
 width of order $2\Gamma$ as predicted from strong-coupling results\cite{Hewson}. $c_2\approx0.3$ to $0.4$ is sufficient to ensure this. With $c_2$ fixed for all interactions, $c_1$ is chosen such that the spectral functions essentially recover the Friedel sum rule, namely $A(0)\simeq1/\pi\Gamma$. 
 {In a simple Anderson model the Friedel sum rule and the parameter $\Gamma$ which governs the width of the Hubbard bands for strong interactions are both readily available. For more complex Anderson-like models information via 
 a Hartree-Fock calculation in the limit of large $U$ for the Hubbard bandwidth and generalizations of the Friedel sum rule should still be obtainable. This implies a general applicability of the proposed broadening scheme.
 
 \begin{figure}[t]
\centering
\includegraphics[width=85mm,height=75mm]{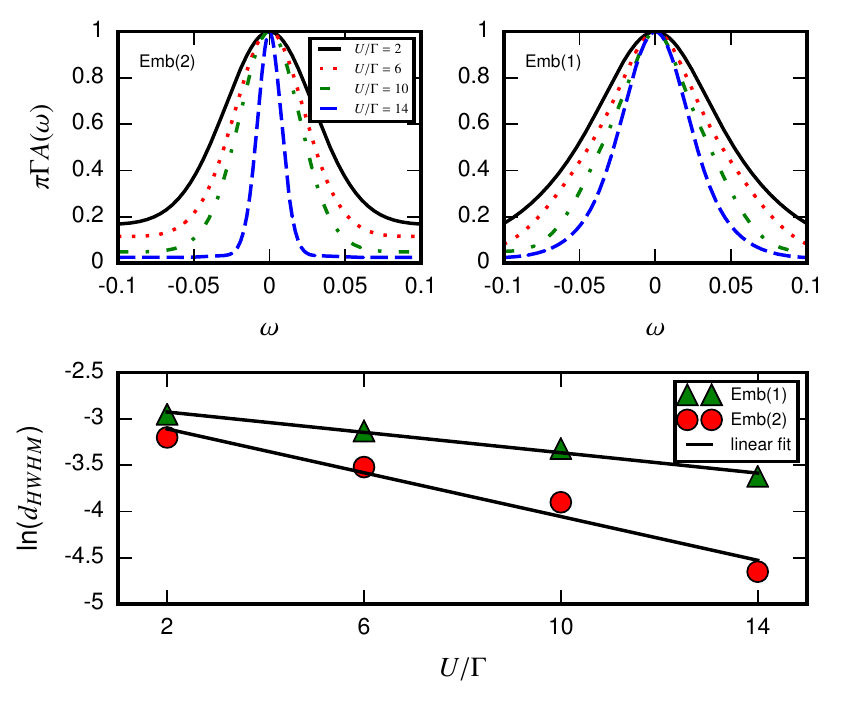}
\caption{\label{fig:kondo_width} Magnified view of the region near $\omega=0$ for the spectral functions considered in Fig.~\ref{fig:siam_dos_log} (top left) and Fig.~\ref{fig:one-site_dos}(top right) 
and the half-width at half-maximum, $d_{HWHM}$, for the different interactions showing the exponential narrowing of the Kondo resonance with increasing $U$ (bottom).}
\end{figure}
 
 Fig.~\ref{fig:siam_dos_log} shows the transfer of spectral weight from low to high-energies at intermediate to strong interactions which marks the canonical distinct features of the SIAM at strong coupling and $T=0$, namely the upper and lower Hubbard satellites, and a zero-frequency peak, the Abrikosov-Suhl or Kondo resonance. 
 A close examination of Fig.~\ref{fig:siam_dos_log} shows that at weak interactions, the two-site embedding method (Emb(2)) overestimates the transfer of spectral weight from the Kondo resonance to the Hubbard bands. 
 However, this is mostly a broadening artifact as the parameter $c_2$, which was optimized to resolve the Hubbard bands accurately, now affects the low energy resonance. A larger value of $c_2$ would remedy this defect. However, this would make the proposed broadening scheme inconsistent. 
 For strong interactions ($U/\Gamma=14$), the peaks of the Hubbard bands are located at $\omega\approx \pm U/2$ which would appear to produce an improvement over the Chebyshev-MPS results. 
 The bottom right panel of Fig.~\ref{fig:siam_dos_log} shows the spectral function obtained with a three-site fragment, Emb(3), using the impurity, the site adjacent to the impurity and the terminal site of the Wilson chain as part 
 of the fragment. 
 The results show that the position of the Hubbard band peaks converge to $\omega = \pm U/2$ for high values of $U$ with increasing fragment size, $n$.

 Fig.~\ref{fig:one-site_dos} shows single-site embedding results for the symmetric SIAM with a linearly discretized conduction band where 299 conduction band levels are uniformly distributed in $\omega \in [-D,D]$. We have considered the wide-band limit analogous to previous two-site embedding calculations shown in Fig.\ref{fig:siam_dos_log} with $\Gamma=0.05$ and $D=1$. 
 With increasing interactions, 
 the narrowing of the Kondo resonance and the appearance of the Hubbard bands is observed. We have used the same broadening scheme to convolute the delta peaks in Fig.~\ref{fig:siam_dos_log} and Fig.~\ref{fig:one-site_dos}. 
 For strong interactions ($U/\Gamma=14$), the Hubbard band peak is located at slightly higher energies than $\pm U/2$.
 For the intermediate coupling regime  ($U/\Gamma=6$), the Hubbard bands seem to be slightly overdeveloped compared to Emb(1) in Fig.~\ref{fig:siam_dos_log} and MPS results. 
 For weak interactions ($U/\Gamma=2$), however, the Emb(1) spectra appears to be more accurate than Emb(2).
 
 The width of the Kondo resonance decreases exponentially with increasing interaction. This trend is captured by both Emb(1) and Emb(2) as seen in Fig.~\ref{fig:kondo_width} 
 where the region near $\omega=0$ is combined for all four panels of Fig.~\ref{fig:siam_dos_log} and Fig.~\ref{fig:one-site_dos} and the half-width at half-maximum, $d_{HWHM}$ is plotted on a log scale for the values of interaction considered in Fig.~\ref{fig:siam_dos_log} and Fig.~\ref{fig:one-site_dos}. 
 For Emb(2), the width of the Kondo peak is somewhat overestimated in the intermediate-coupling regime. For Emb(1), the narrowing of the Kondo resonance is not as sharp as the two-site case but still follows the correct trend.
 For all the spectral functions, the Friedel sum rule is nearly trivially obeyed with deviations below $1\%$ with an appropriate choice of $c_1$. This condition, being imposed, does not automatically reflect the accuracy of the low-energy features obtained from the embedding method.

\section{Conclusion}\label{conclusion}
In this work we outline and investigate a very simple embedding approach for calculating the spectral properties of the single impurity Anderson model. 
The method essentially combines features from density matrix embedding theory (DMET) with discretization and broadening schemes adopted from numerical renormalization
group (NRG) approaches.  We advocate a `single-shot' embedding scheme whereby no self-consistency is required, even for cases where multiple bath sites are 
included in the embedding.  The approach provides results compatible with much more advanced DMRG-based algorithms at a fraction of the numerical complexity and 
expense.

The approach discussed in this work should find useful application as an impurity solver for DMFT, as a prior for analytical continuation of imaginary-time quantum
Monte Carlo, and for investigation of the spectral features of more complex impurity problems in their own right.  With respect to the latter, it should be noted here that much more complex
impurity problems, such as those with multiple orbitals and coupled sites are amenable to the method discussed here with limited additional expense.  Future work will
be devoted to study of both more complex impurity problems such as multi-orbital cases that emerge in DMFT as well as testing our approach on the more subtle and
intricate two-impurity Anderson model.

\begin{figure}[t]
\centering
\includegraphics[width=85mm,height=65mm]{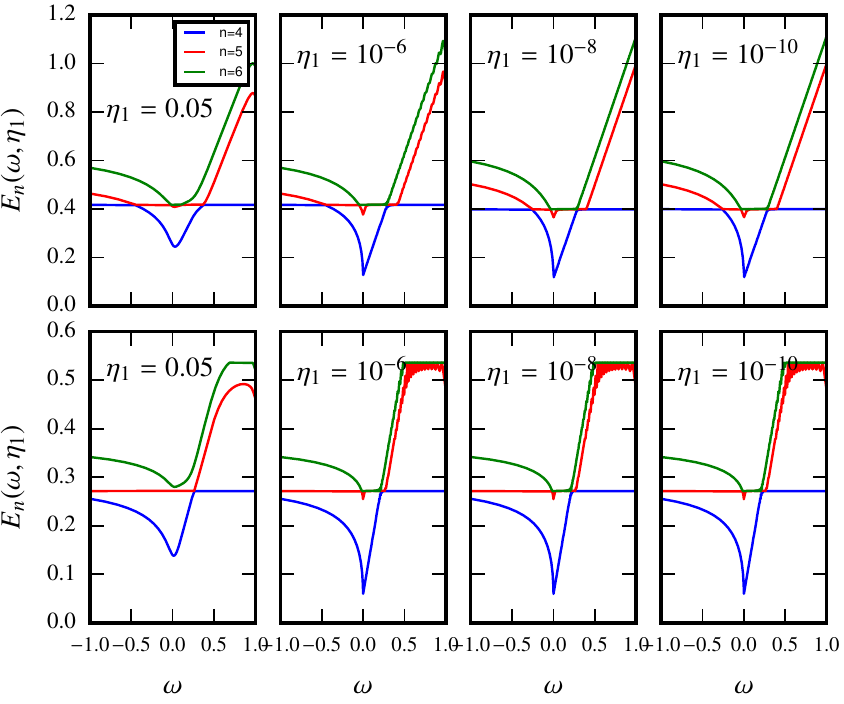}
\caption{\label{fig:app_fig1} $E'_n(\omega, \eta_1)$ for $n=4,5,6$ and different values of $\eta_1$ at $U/\Gamma = 10$. Top row refers to Emb(1) and bottom row to Emb(2).}
\end{figure}
While this paper was being written we became aware of a related work by Kretchmer and Chan\cite{td_dmet}. This works investigates non-equilibrium transport in the Anderson impurity 
model while we focus, with a slightly different methodology, on the equilibrium spectral properties.

\section{Acknowledgements}
SM thanks Garnet K. Chan for useful discussions. DRR acknowledges funding from NSF CHE-1464802.

\section{Appendix}
\subsection{Choice of $\eta_1$ to ensure convergence of delta peaks}\label{app: eta1}
\begin{figure}[t]
\centering
\includegraphics[width=85mm,height=65mm]{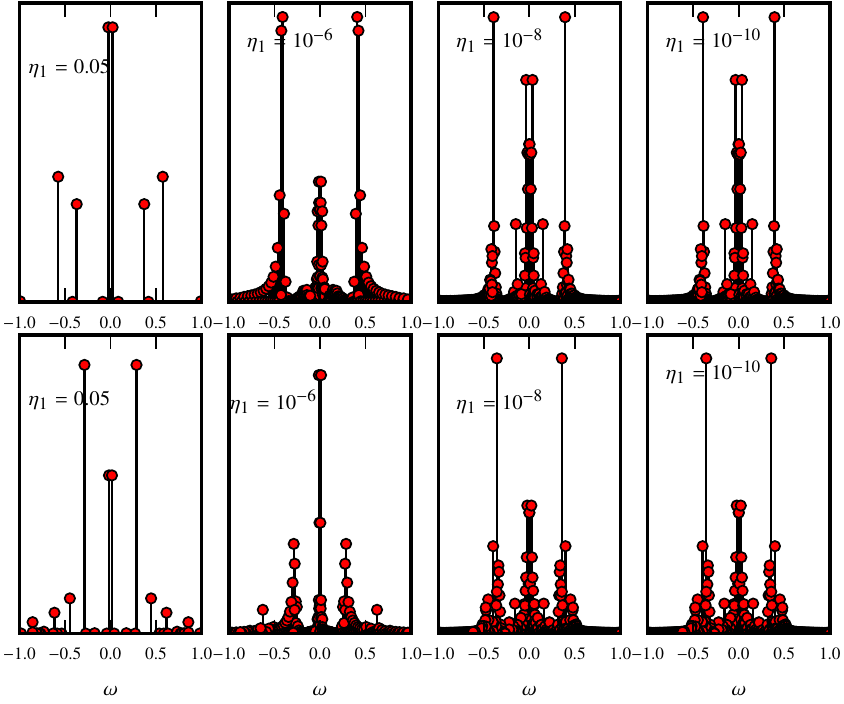}
\caption{\label{fig:app_fig2} Peak spectrum for different values of $\eta_1$ at $U/\Gamma = 10$. Top row refers to Emb(1) and bottom row to Emb(2).}
\end{figure}
In this short appendix we provide more details concerning the determination of broadening parameters and the sensitivity of the spectra to these parameters. 
The value of $\eta_1$ determines the number of delta peaks in the peak spectrum obtained from the Lehmann representation of the single-particle Green's function, see Eq.~\ref{eq:gf_pre_peak} and Eq.~\ref{eq:lehmann}. 
Fig.~\ref{fig:app_fig1} shows the behavior of $E_{n=4,5,6}(\omega, \eta_1)$ for different values of $\eta_1$. For Emb(1), $E_n(\omega, \eta_1)$ is not sensitive below $\eta_1=10^{-8}$ while, for Emb(2), $\eta_1=10^{-6}$ ensures convergence 
for $E_n(\omega, \eta_1)$. The choice of $\eta_1$ is primarily determined by examining the peak spectrum in Fig.~\ref{fig:app_fig2}. The structure and the number of peaks in Fig.~\ref{fig:app_fig2} converges for $\eta_1=10^{-8}$ which is the 
value used throughout this work.




\providecommand{\noopsort}[1]{}\providecommand{\singleletter}[1]{#1}%

\end{document}